\documentclass[showpacs,showkeys,floatfix]{revtex4}
\usepackage[dvips]{graphicx}
\usepackage{epsfig}

\usepackage[cp1251]{inputenc}
\usepackage[english]{babel}

\begin{document}

\title{Dynamic and spectral mixing in nanosystems.}
 
\author{V.A.Benderskii} 
\affiliation {Institute of Problems of Chemical Physics, RAS \\ 142432 Moscow
Region, Chernogolovka, Russia} 

\author{E. I. Kats} \affiliation{Laue-Langevin Institute, 
Grenoble, France} 
\affiliation{L. D. Landau Institute for Theoretical Physics, RAS, 
Moscow, Russia}

\date{\today}

\begin{abstract}

In the framework of simple spin-boson Hamiltonian we study an interplay 
between dynamic and spectral roots to stochastic-like behavior. The Hamiltonian describes an initial vibrational state coupled to discrete dense spectrum reservoir. The reservoir states are formed by three sequences with rationally independent periodicities $1\, ;\, 1 \pm \delta $ typical for vibrational states in many nanosize systems (e.g., large molecules containing $C H_2$ fragment chains, or carbon nanotubes). We show that quantum evolution of the system is determined by a dimensionless parameter $\delta \, \Gamma $, where $\Gamma $ is characteristic number of the reservoir states relevant for the initial vibrational level dynamics. When $\delta \, \Gamma > 1$ spectral chaos destroys recurrence cycles and the system state evolution is stochastic-like. In the opposite limit $\delta \, \Gamma < 1$ dynamics is regular up to the critical recurrence cycle $k_c$ and for larger $k > k_c$ dynamic mixing leads to quasi-stochastic time evolution. Our semi-quantitative analytic results are confirmed by numerical solution of the equation of motion. We anticipate that both kinds of stochastic-like behavior (namely, due to spectral mixing and recurrence cycle dynamic  mixing) can be observed by femtosecond spectroscopy methods in nanosystems in the spectral window $10^{11} \, -\, 10^{13} \, s^{-1} $.

\end{abstract}


\maketitle

Quantum dynamics for an initial vibrational level $\epsilon _s^0$ coupled to a set of discrete dense levels $\epsilon _n^0$ can be described in a framework of so-called spin-boson Hamiltonian \cite{CL83}, \cite{LC87}

\begin{equation}\label{n1}
H = \epsilon _s^0 b_s^+b_s + \sum _{n}\epsilon _n^0 b_n^+b_n + \sum _n (b_n^+b_s + b_n b_s^+)
\, ,
\end{equation}
where $b_s^+(b_s)$ is initial level creation (annihilation) operator (i.e., excitation of the initial vibrational level from the system 
ground state, which is assumed so deep that its influence on the system dynamics can be neglected), and $b_n^+(b_n)$ are similar operators for the discrete reservoir levels. $C_n$ stands for the coupling matrix elements. For this Hamiltonian time dependent wave function can be presented in a series over full orthogonal set of wave functions of the unperturbed (uncoupled) initial and reservoir states with time dependent coefficients (amplitudes) $a_s(t)$

\begin{equation}\label{n2}
a_s(t) = \sum _{n}\left \{\left (\frac{d F}{dE}\right )^{-1} \exp (- i E t)\right \}_{E = \epsilon _n}
\, ,
\end{equation}
where $F(E) = 0$ is a secular equation to find the eigenstates $\epsilon _n$ of the Hamiltonian (\ref{n1}).

Common wisdom \cite{GR93}, \cite{HA01}, \cite{ME04}, \cite{PW07} claims that stochastic-like dynamics is a feature typical for random matrix Hamiltonians, with random eigenstate spectra. However for a system with discrete dense spectrum (e.g., vibrational states in nano-particles or medium size molecules) there is another dynamic root to quasi-stochastic behavior. Indeed for such a system with discrete spectrum, dynamic evolution is represented by periodically repeating steps (recurrence cycles). When time is going on, the initial vibrational level population oscillates faster and faster and corresponding response signals become broader. Eventually at a certain critical cycle number $k_c$, the cycles are overlapped in time. Then for any finite accuracy of time or frequency measurements, time evolution looks as irregular, quasi-stochastic, indistinguishable from truly chaotic behavior. It was demonstrated recently \cite{BF07}, \cite{BG09}, \cite{BK09} for the simplest version of the spin-boson Hamiltonian (\ref{n1}) with $\epsilon _n^0 \equiv n$,
and $C_n \equiv C$ (so-called Zwanzig approximation \cite{ZW60}).
Here (and in what follows) we utilize a characteristic reservoir spacing as energy unit, and $n=0$ is the reservoir level coinciding with $\epsilon _s$. 
It is worth noting that this dynamic mixing occurs in a single Hamiltonian system (not for an ensemble of equivalent systems).

Since there are mentioned above two roots to stochastic-like time evolution, two natural questions arise, namely, (i) how the both mechanisms are interrelated, and (ii) whether they are independent ones. To answer these questions is a purpose of our presentation. We study quantum dynamics for a version of spin-boson Hamiltonian (\ref{n1}), where the both ingredients yielding to stochastic time evolution (spectral and dynamic) 
may be presented simultaneously and can be tuned by model Hamiltonian parameters. We assume that the reservoir discrete bare spectrum in the Hamiltonian (\ref{n1}) is formed by three equidistant sequences with different periods
\begin{equation}\label{n3}
\epsilon _n^0 = \pm 3 n \, ;\, \epsilon _1^0 = \pm (1-\delta )(3n + 1)\, ;\, \epsilon _2^0(n) = \pm (1+ \delta )(3n +2)
\, ,
\end{equation}
where spectral shift dimensionless parameter $\delta \leq 1/2$. It is worth noting that such kind of triplet structures are quite frequently observed in the spectra of many molecular systems, forming notorious Snyder sequences \cite{SN60} (see also more recent publications on high resolution  spectroscopy of nanotubes \cite{IS03}, phospholipid molecules \cite{RO04}, \cite{TH10} and fullerenes \cite{CH06}).
When the total number of levels $N \to \infty $ and the sequence periods $1\, ,\, 1 \pm \delta $ are rationally independent, the spectrum of the reservoir becomes mixed. 
In own turn, according to the ergodic properties \cite{SI76}, \cite{GR06} chaotic behavior  of an ensemble of systems (random eigen-value distribution) holds if its individual system is mixing. 
That is why we term our case as exhibiting of stochastic-like behavior (cf. with definition \cite{LU04}).  

In more practical terms mixing phenomenon can be related to reservoir level ordering.
Indeed, level ordering and interlevel spacings depend on a cycle number. In a zero cycle the spectrum is formed by triplets in the following sequence
\begin{equation}\label{n4a}
\epsilon _2(n-1)  < \epsilon _0(n) < \epsilon _1(n) < \epsilon _2(n)
\, .
\end{equation}
When $n$ increases the quantities $p_1^{(0)} = \epsilon _0(n) - \epsilon _2(n-1)$ and  $p_2^{(0)} = \epsilon _1(n) - \epsilon _0(n)$ become smaller and fill with a step $3\delta $ intervals $[0\, , \, 1 \pm \delta ]$. Level splitting between neighboring triplets $p_3^{(0)} = \epsilon _1(n) - \epsilon _0(n)$
increases with a step $6\delta $ approaching to 3. In the next cycle $k=1$ the level ordering is different
\begin{equation}\label{n4b}
\epsilon _1(n) < \epsilon _0(n) < \epsilon _2(n-1) < \epsilon _1(n+1)
\, ,
\end{equation}
and the quantities $p_1^{(1)}(n) = \epsilon _2(n-1) - \epsilon _0(n)$ and $p_2^{(1)}(n) = \epsilon _0(n) - \epsilon _1(n)$ are increased with the same as in the $k=0$ cycle, step $3\delta $ up to a limit value 3/2. One can check that in the next even cycles the splittings $p_1^{(2k)}$ and $p_2^{(2k)}$ decrease, whereas $p_3^{(2k)}$ increases. In the odd cycles the opposite variation of the splittings hold ($p_1$ and $p_2$ increase, and $p_3$ decreases). Eventually levels from different triplets fill in the $k \to \infty $ limit almost uniformly and densely 
the interval $[0\, ,\, 6\delta ]$ with odd-even alternations with the recurrence cycle number. 
By rather lengthy and boring but straitforward calculations we find that for cycle number $k \geq 1$, the variables $p_1$ and $p_2$ fill uniformly the intervals
\begin{equation}\label{n5}
\left [a_k^{(1)}\, ,\, \frac{3}{2} - a_{k+1}^{(1)}\right ] \, ; \, \left [a_k^{(2)}\, ,\, \frac{3}{2} - a_{k+1}^{(2)}\right ]
\, ,
\end{equation}
with the different order of $p_1$ and $p_2$ in even and odd cycles.
The $p_3$ fills the interval
\begin{equation}\label{n6}
\left [3 - a_k^{(3)}\, ,\, a_{k+1}^{(3)}\right ]
\, .
\end{equation}
In own turn the above intervals limits $a_k^{(j)}$ with $j = 1 , 2 , 3$ satisfy
the following relations
\begin{equation}\label{n7a}
a_{k}^{(1 , 2)} = \left \{\frac{1}{6\delta }(3k + 2) \pm \frac{1}{3} \right \} \, ;\, a_{k}^{(3)} = \left \{\frac{1}{6\delta }(3k+2) - \frac{1}{2}\right \}
\, ,
\end{equation}
where $\{X\}$ stands for the fractional part of $X$.
At $k \gg 1$ the (\ref{n7a}) can be generated approximately by so-called fractional part recurrence relation
\begin{equation}\label{n7}
a_{k+1}^{(j)} \simeq \left \{\frac{1}{2\delta }\, a_k^{(j)}\right \}
\, .
\end{equation}
As it is known \cite{ZA85} the fractional part transformation is the mixing one (i.e., stochastic) with characteristic correlation time (in our notations) $\propto \ln (1/\delta )$. Thus our model reservoir spectrum is the reservoir with spectral mixing. This phenomenon can be formulated as the mixing of $p_j$ parameters within the interval $[0\, ,\, 3]$. The interlevel spacing distribution density is approximately (over the parameter $1/N$) a constant $\simeq 5/9$ in the interval $[0\, ,\, 3/2]$ and another constant
$\simeq 1/9$ in the interval $[3/2\, ,\, 3]$.
Without going to more subtle mathematical details of random sequences we calculate numerically the distribution function $\rho (\epsilon )$ for our model spectrum. When the total number $N$ of the reservoir levels increases the distribution approaches to that with two uniformly distributed parts (see the fig. 1, where we show also the widely used in the literature Wigner distribution (see e.g., \cite{ME04}, \cite{PW07}, and also
\cite{WI67}) which holds for orthogonal Gaussian random matrices \cite{ME04}).
\begin{figure}[htb]
   \centering
        \includegraphics[width=0.5\textwidth]{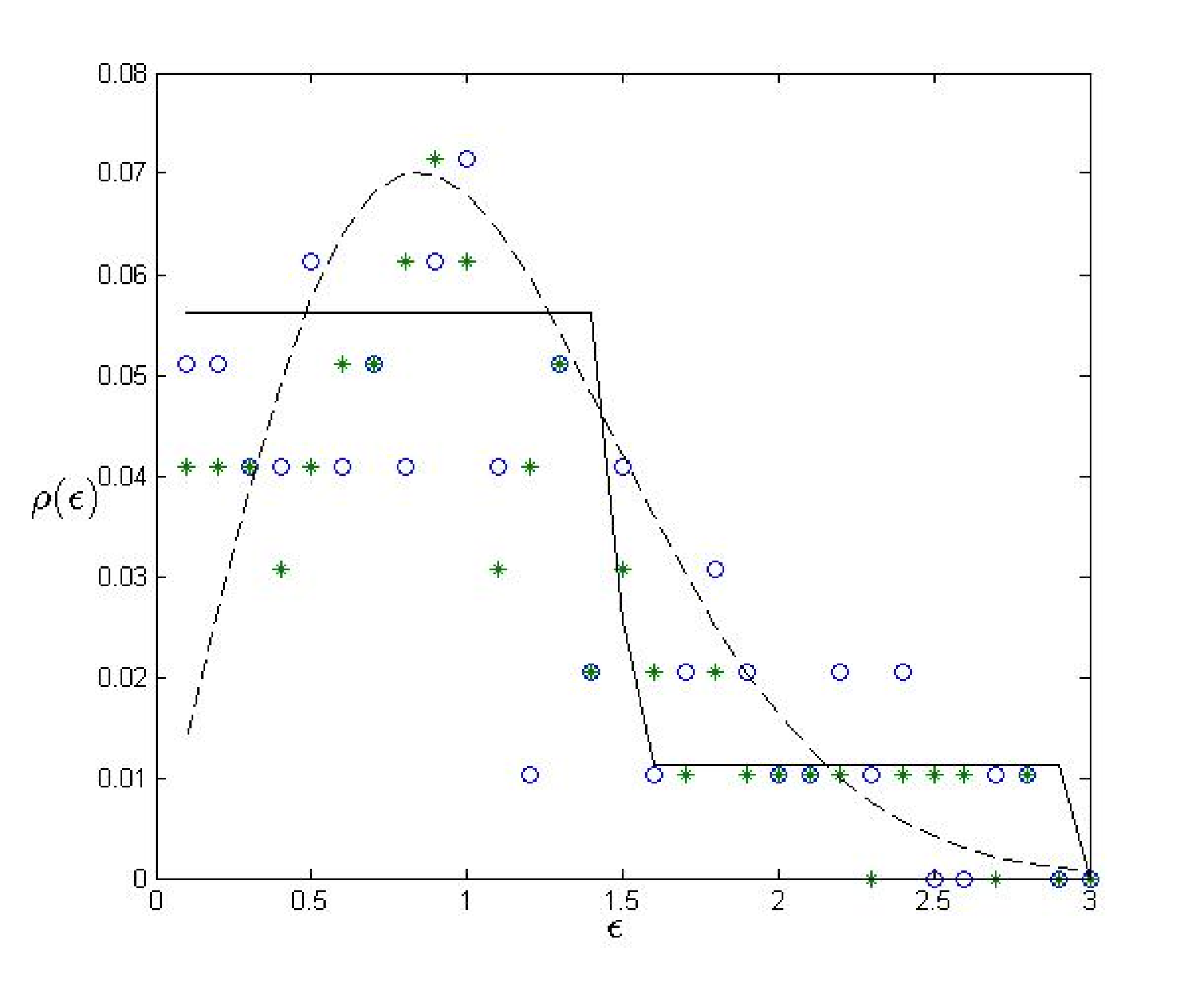}
    \caption{Interlevel spacing distribution density $\rho (\epsilon )$ for the reservoir (\ref{n3}) with 3 rationally 
   independent periods ($\delta = 0.049$). Empty circles ($C^2=0$), stars ($C^2=1$), and filled circles (both distribution densities coincide) are %numerical computation performed for 166 levels (10 cycles). Solid line - our theoretical prediction with level 
   density 5/9 in the interval $[0\, , \, 3/2]$ and another constant 1/9 in the interval $[3/2\, ,\, 3]$. 
   Wigner distribution is shown for comparison by dashed line.}
    \label{fig:1}
\end{figure}

Let us come back to the secular equation to find the eigenstates
\begin{eqnarray}
\label{n8}
F(\epsilon ) = \epsilon - \sum_{n=-N}^{n = + N}\left (\frac{C_{3n}^2}{\epsilon -\epsilon _{3n}^0}
+ \frac{C^2_{3n+1}}{\epsilon - \epsilon _{3n+1}^0} 
+ \frac{C_{3n+2}^2}{\epsilon -\epsilon _{3n+2}^0}\right ) = 0
\, .
\end{eqnarray}
For the ease of algebra (if needed this approximation can be relaxed) we assume following Zwanzig \cite{ZW60} approximation
$$
C_{3n} = C_{3n+1} = C_{3n+2} \equiv C
$$
and in this case the series entering (\ref{n8}) can be written explicitly in terms of trigonometric functions
\begin{equation}\label{n9}
F(\epsilon ) = \epsilon - \frac{\Gamma }{3}\left [\cot \left(\frac{\pi \epsilon }{3}\right ) + \cot \left (\frac{\pi \epsilon }{3(1-\delta )}- \frac{\pi }{3}\right ) + \cot \left (\frac{\pi \epsilon }{3(1+\delta )} - \frac{2\pi }{3} \right ) \right ] = 0
\, .
\end{equation}
Here $\Gamma = \pi C^2$ characterizes the window of the reservoir states, where at $\delta = 0$ the coupling to the initial vibrational level $\epsilon _s =0$ contributes essentially to system time evolution (the reservoir levels with the quantum numbers $\leq \Gamma $ govern the system dynamics).

When $\delta = 0$ our model is reduced to a single equidistant reservoir Zwanzig model, and in this case dynamic stochastic-like behavior (due to recurrence cycle mixing) occurs \cite{BF07}, \cite{BK09} for the cycle number $k > k_c^0 = \pi \Gamma $. If $\delta \neq 0$ analysis is more involved. One can check by tedious but direct calculations that the trigonometric equation (\ref{n9}) has one and only one root in each interval between the values of the bare energies $\epsilon _0^0(n)\, ,\, \epsilon _1^0(n)$, and $\epsilon _2^0(n)$ given by the expressions (\ref{n3}). Armed with this knowledge the formal solution (\ref{n2}) for the initial vibrational level amplitude $a_s(t)$ can be represented as a finite series over the partial recurrence cycle amplitudes $a_s^{(k)}(\tau _k)$ (where $k$ is confined within the interval $0 \leq k \leq [t/(2\pi )]$ and $[y]$ means an integer part of $y$) calculated in the local cycle time $\tau _k = t - 2\pi k$. 
In own turn $a_s^{(k)}(\tau _k)$ can be presented in terms of generalized Lommel and Laguerre polynomials \cite{BE53}. To avoid this lengthy and cumbersome mathematics in this publication we restrict ourselves to the only small $\delta \ll 1$ limit, when the secular equation is reduced to
\begin{equation}\label{n10}
\epsilon - \Gamma \cot (\pi \epsilon )[1 + R(\epsilon \, ,\, \delta )] = 0
\, ,
\end{equation}
with
\begin{eqnarray}
\label{n11}
R(\epsilon \, ,\, \delta ) = \left (1 + q\frac{\sin (\pi \epsilon /3)}{\sin (\pi \epsilon )}\right )^{-1}\left [q\left (\frac{\sin (\pi \epsilon /3)}{\sin (\pi \epsilon )} - \frac{\cos (\pi \epsilon /3)}{3\cos (\pi \epsilon )}\right )
- 2 \delta \frac{\sin (\pi \epsilon /3)}{\cos (\pi \epsilon )}(\sqrt {3} +q) \right ]
\, ,
\end{eqnarray}
where
\begin{equation}\label{n12}
q = 1 + 4 \sin(\pi \delta \epsilon )\cos (\pi \delta \epsilon + \pi /3)
\, .
\end{equation}
All qualitative features of the quantum time evolution for our model can be seen in this limit $\delta \ll 1$. Since $F(\epsilon )$ depends on $\pi \epsilon /3$ in each recurrence cycle $k$ its amplitude $a_s^{(k)}$ has not only the main Loschmidt echo signal but also two sattelites shifted with respect to the main signal by $2\pi /3$ and $4\pi /3$ (see fig. 2 where we plot initial vibrational level population time evolution). Intensities of the sattelites are proportional to $\delta $, and when the cycle number $k$ increases the main echo signal and its sattelites acquire a fine internal structure. The number of  components in the fine structure of the echo signals increases with $k$ and starting with a certain threshold $k_c$ start to overlap. For $\delta \ll 1$
\begin{equation}\label{n13}
k_c^{-1} = (\pi \Gamma )^{-1} + 3 \delta 
\, .
\end{equation}
At $k> k_c$ the triplets formed by the main echo signal and sattelite triplets are strongly mixed and time evolution becomes chaotic. In the fig. 3 we computed $k_c(\delta )$. It is clear that the numerical results are in a good agreement with our approximated formula (\ref{n13}).
\begin{figure}[htb]
   \centering
      \includegraphics[width=0.5\textwidth]{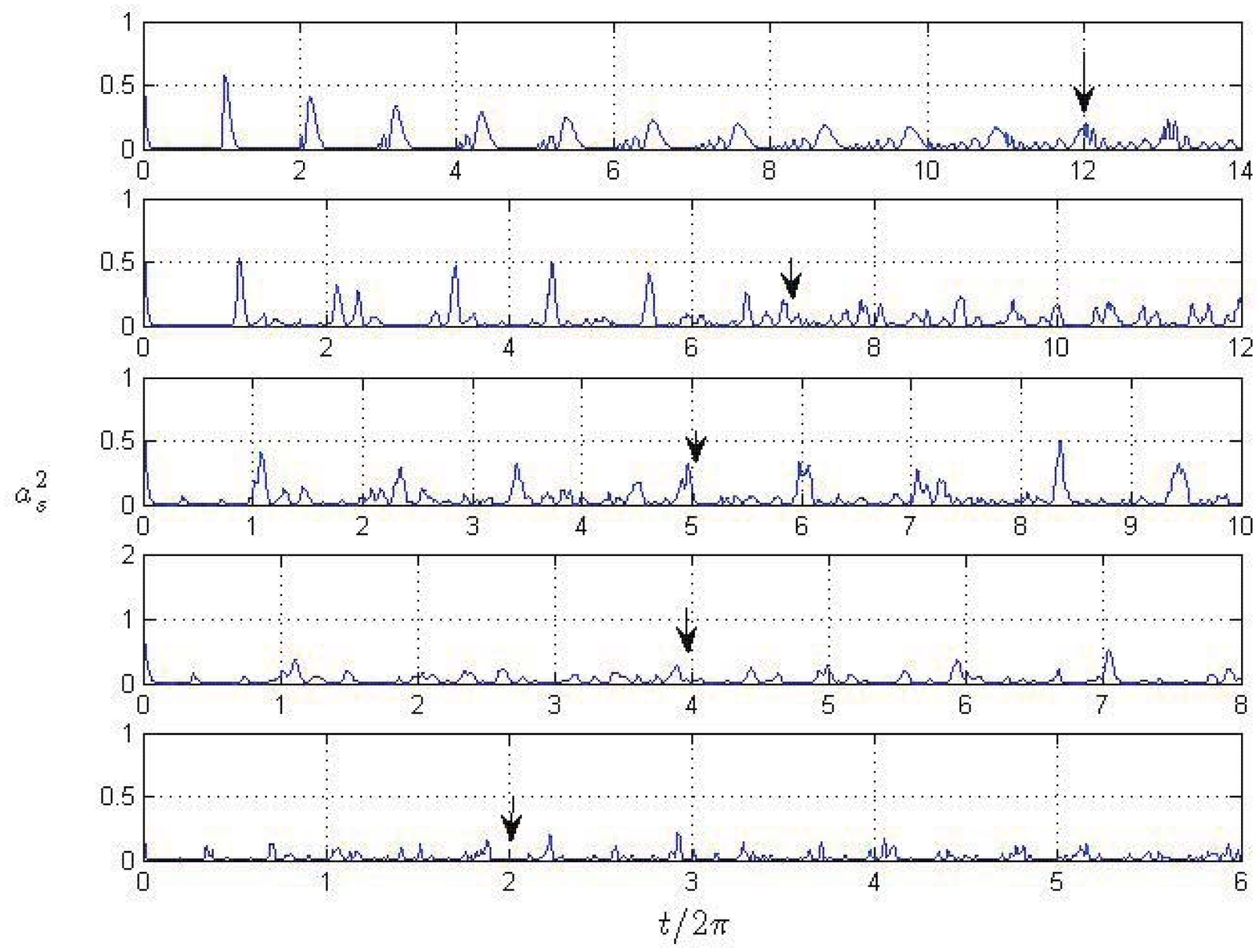}
   \caption{Initial vibrational level population dynamics ($C^2=1$). From the top to the 
bottom: $\delta = 0 \, ;\, \delta = 0.019 \, ;\, \delta =  0.049 \,  ;\, \delta = 0.079 \, ; \, \delta = 0.12 $. 
Critical cycle numbers $k_c$  are indicated by arrows.}
    \label{fig:2}
\end{figure}
This formula (\ref{n13}) is our main result in this paper. It shows how the spectral source of the chaotic behavior ($\delta $ in the (\ref{n13})) interplays with the dynamic source of stochastic-like time evolution ($\Gamma $ in the (\ref{n13})). Namely, $k_c(\delta )$ is determined by the condition that the triplets from the cycle $k_c$ are mixed with those from the next cycle. The spectral chaos contribution ($\delta \neq 0$) reduces considerably the number of cycles with regular dynamics. For $\delta = 0.12 $ (still $\delta \ll 1$) the intensities of sattelites and the main echo component are comparable and the regular triplet structure is completely broken after 2 initial cycles. We illustrate this in the fig. 2, where we show numerically calculated $|a_s(t)|^2$ (i.e., the initial vibrational level population) for $\delta = 0\, ,\, 0.019\, ,\, 0.049\, ,\, 0.079\, ,\, 0.12 $
(a bit bizarre numerical values of $\delta $ are chosen to get better approximation for rationally independent sequences of the reservoir levels).
\begin{figure}[htb]
   \centering
      \includegraphics[width=0.5\textwidth]{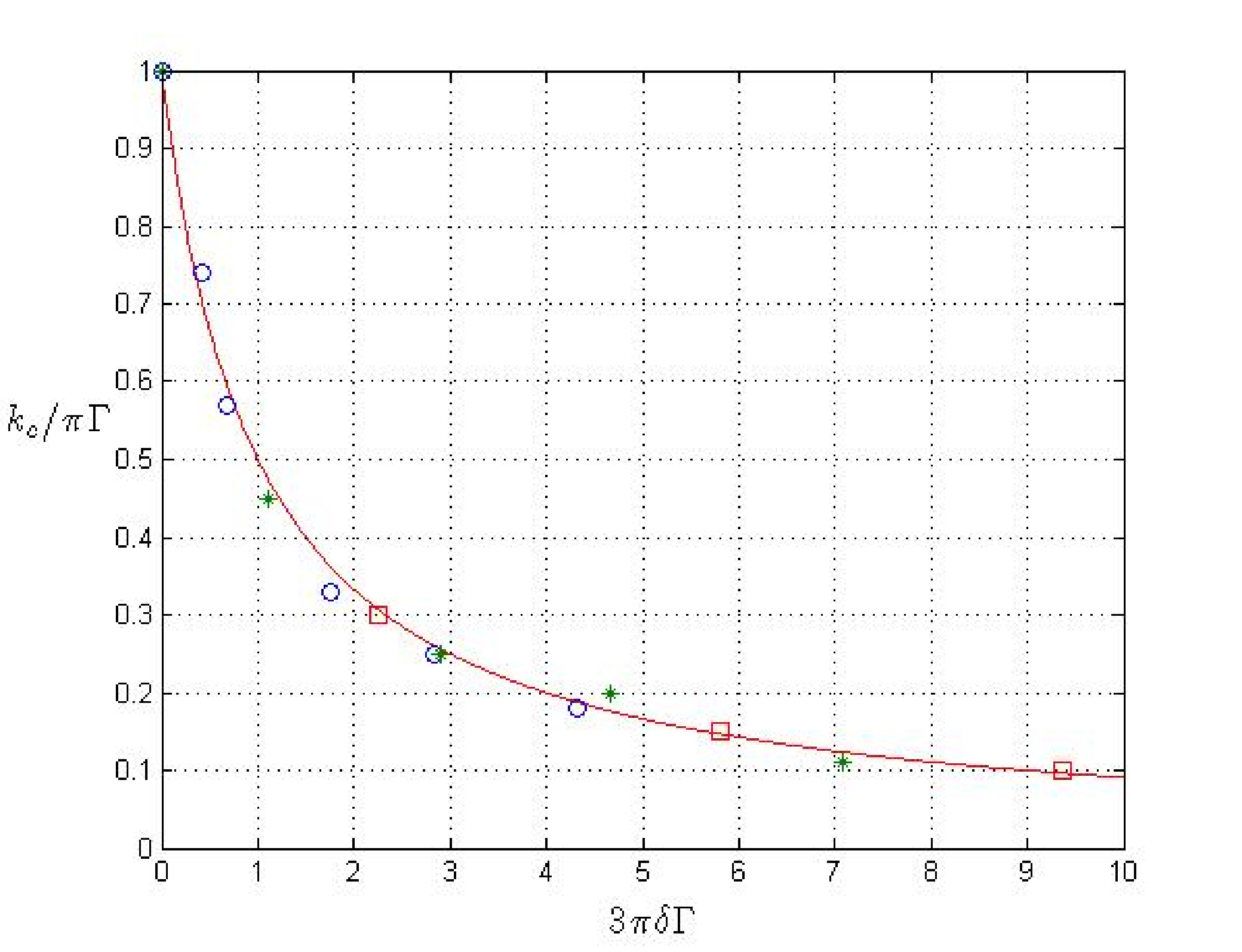}
    \caption{Critical cycle number dependence on $\delta $: $C^2=1$ - open circles; $C^2 = 2$ - stars; $C^2 =4$ - squares.}
   \label{fig:3}
\end{figure}

Thus we arrive at the general conclusion. Dynamic, stochastic-like time evolution is determined by the local spectrum characteristics (mean interlevel spacing) and occurs only for sufficiently large recurrence cycle number. On the contrary spectral truly chaotic dynamics is governed by the global spectrum structure (e.g., in our spin-boson model Hamiltonian (\ref{n1}) by the reservoir triplet structure (\ref{n3})). 
It is instructive also to show contributions of the reservoir states into the initial vibrational level wave function amplitude $a_s(t)$. According to the expression (\ref{n2}) we computed the quantity $A_n \equiv (d F/dE)^{-1}|_{E = \epsilon _n}$ and plotted the results in the fig. 4.
These quantities $A_n$ also manifest themselves simultaneous effects of dynamic and spectral mixing. For the pure Zwanzig model (no spectral mixing, i.e., $\delta =0$), the coefficients $A_n$ decrease monotonously with $\epsilon _n$. However, upon increasing $\delta $ (spectral mixing),
$A_n$ distribution becomes more and more irregular.

\begin{figure}[htb]
   \centering
      \includegraphics[width=0.5\textwidth]{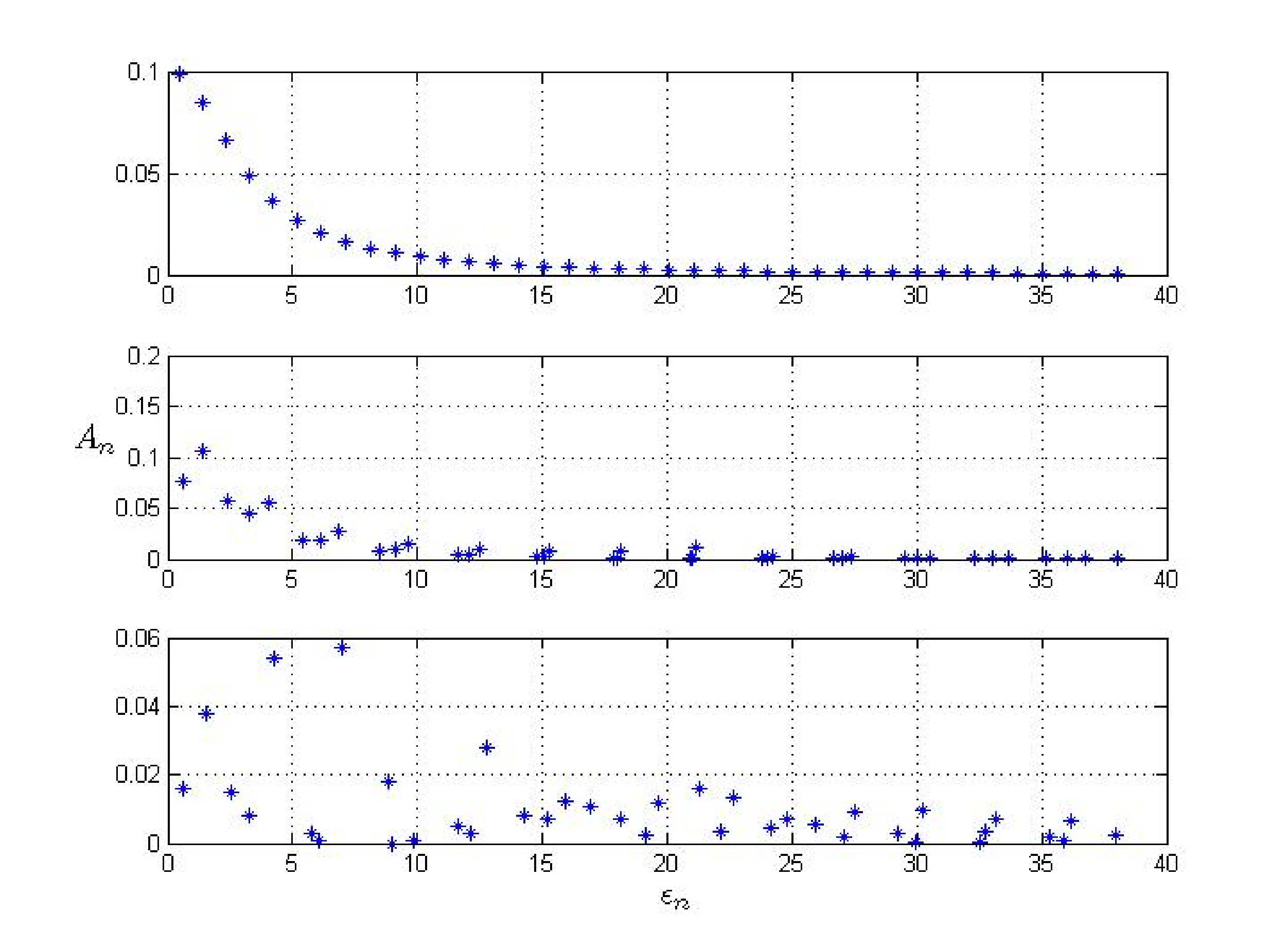}
    \caption{Amplitudes $A_n \equiv \left (\partial F/\partial E \right )^{-1}_{E=\epsilon _n}$ (\ref{n2}) of the reservoir states contributing into 
   the initial vibrational level evolution. From the top to the bottom: $\delta =0 \, ;\, \delta = 0.049 \, ;\, \delta = 0.12 $, and  $C^2=1$.}
    \label{fig:4}
\end{figure}

Our motivation in this work is not a pure curiosity. As a matter of fact quantum dynamics of various systems (ranging from 
relatively small molecules up to large photochromic molecules and their protein
complexes or molecules confined near interfaces \cite{BE02} (see also \cite{FE03})
is an active area of 
experimental researches. The femtosecond spectroscopy data (which allow to study time evolution of one  initially prepared
by optical pumping state) manifest variety of possible regimes including not only weakly damped more or less regular oscillations
but also very irregular long time behavior with a number of peaks corresponding to a partial recovering of the initial state
population. 
This generic feature is omnipresent in the systems with complex and irregular vibrational relaxation.
The triplet model investigated in this paper reflects the spirit of minimalist approaches, in that it is simple yet based on a 
physical principle.
Understanding all its limitations, we nevertheless hope that our crude theory captures the essential elements of vibrational relaxation in 
nano-systems.

\newpage

\centerline{Figure captions}

Fig. 1

Interlevel spacing distribution density $\rho (\epsilon )$ for the reservoir (\ref{n3}) with 3 rationally independent periods ($\delta = 0.049$). Empty circles ($C^2=0$), stars ($C^2=1$), and filled circles (both distribution densities coincide) are numerical computation performed for 166 levels (10 cycles). Solid line - our theoretical prediction: $\rho (\epsilon ) \simeq 5/9$ in the interval $[0\, , \, 3/2]$ and 
$\rho (\epsilon ) \simeq 1/9$ in the interval $[3/2\, ,\, 3]$. Wigner distribution is shown for comparison by dashed line.

Fig. 2

Initial vibrational level population dynamics ($C^2=1$). From the top to the bottom: $\delta = 0 \, ;\, \delta = 0.019 \, ;\, \delta = 0.049 \,  ;\, \delta = 0.079 \, ; \, \delta = 0.12 $. Critical cycle numbers $k_c$  are indicated by arrows.

Fig. 3 

Critical cycle number dependence on $\delta $: $C^2=1$ - open circles; $C^2 = 2$ - stars; $C^2 =4$ - squares.

Fig. 4

Amplitudes $A_n \equiv \left (\partial F/\partial E \right )^{-1}_{E=\epsilon _n}$ (\ref{n2}) of the reservoir states contributing into the initial vibrational level evolution. From the top to the bottom: $\delta =0 \, ;\, \delta = 0.049 \, ;\, \delta = 0.12 $, and $C^2=1$.


\begin{references}

\bibitem{CL83} 
A.O.Caldeira, A.J.Leggett, Ann. Phys., {\bf 149}, 587 (1983).

\bibitem{LC87}
A.J.Leggett, S.Chakravarty, A.T.Dorsey, M.P.A.Fisher, A.Garg, M.Zweger, Rev. Mod. Phys., {\bf 59}, 1 (1987).

\bibitem{GR93} 
P.Grigolini, Quantum Mechanical Irreversibility, World Scientific, Singapore (1993).

\bibitem{HA01} F.Haake, Quantum signature of chaos, 2d Edition, Springer, Berlin (2001).

\bibitem{ME04} M.L.Mehta, Random matrices, 3d Edition, Academic Press, New York (2004).

\bibitem{PW07} T.Papenbrock, H.A.Weldenmuller, Rev. Mod. Phys., {
bf 79}, 997 (2007).


\bibitem{BF07} 
V.A.Benderskii, L.A.Falkovsky, E.I.Kats, JETP Lett., {\bf 86}, 311 (2007).


\bibitem{BG09} V.A.Benderskii, L.N.Gak, E.I.Kats, JETP, {\bf 108}, 160 (2009), and JETP, {\bf 109}, 505 (2009).

\bibitem{BK09} V.A.Benderskii, E.I.Kats, Eur. Phys. J., D, {\bf 54}, 597 (2009).


\bibitem{ZW60} 
R.Zwanzig, Lectures in Theor. Phys., {\bf 3},
106 (1960).


\bibitem{SN60} R.C.Snyder, J.Mol. Spectr., {\bf 4}, 411 (1960).

\bibitem{IS03} T.Ishioka, et al., Spectrochimica Acta A, {\bf 59}, 671 (2003).

\bibitem{RO04} K.R.Rodriguez, et al., J. Chem. Phys., {\bf 121}, 8671 (2004).

\bibitem{CH06} O.P.Charkin, et al., J. Inorg. Chem., {\bf 51}, Suppl.1,  1 (2006).

\bibitem{TH10} J.M.Thomas, Angew. Chem., {\bf 43}, 2606 (2010).

\bibitem{SI76} Ya.G.Sinai, Introduction to ergodic theory, Princeton University Press, Princeton (1976).

\bibitem{GR06} R.M.Gray, Probability, random processes and ergodic properties, Springer, Berlin (2006).

\bibitem{LU04} S.Luzatto, Arxiv, Math. (2004).

\bibitem{ZA85} G.M.Zaslavskii, Chaos in dynamical systems, Harwood, New York (1985).

\bibitem{WI67} E.P.Wigner, SIAM Rev., {\bf 9}, 1 (1967).


\bibitem{BE53} 
H.Bateman, A.Erdelyi,
Higher Transcendental Functions, vol.2, McGraw Hill, New
York (1953).


\bibitem{BE02} 
A.V.Benderskii, K.B.Eisental, J. Phys. Chem., A, {\bf 106}, 7482 (2002).



\bibitem{FE03} 
C.J.Fecko, J.D.Eaves, J.J.Loparo, A.Tokmakoff, P.L.Geissler, Science, {\bf 301}, 1698 (2003).




\end{references}
\end{document}